\documentclass[10pt,conference]{IEEEtran}

\usepackage[numbers]{natbib}
\usepackage{amsmath}
\usepackage{amsfonts}
\usepackage[rounded]{syntax}
\usepackage{listings}
\usepackage{graphicx}
\usepackage{url}
\usepackage{float}
\usepackage{stfloats}
\usepackage{booktabs}
\usepackage[center,nooneline,tight]{subfigure}
\usepackage{listings}

\usepackage{xcolor}
\usepackage{setspace}   
\newcommand{\click}{{\textsc{Cl\makebox[.58\width][c]{1}ck}}}

\definecolor{darkblue}{RGB}{40, 40, 244}

\newcommand{\mypar}[1]{{\bf #1.}}

\newcommand{\eg}{e.g.}
\newcommand{\ie}{i.e.}

\newcommand{\mydsl}{LL}
\newcommand{\sll}{$\Sigma$-\mydsl}

\newcommand{\R}[0]{\mathbb{R}}

\newfloat{Grammar}{ht}{lop}

\begin{document}

\title{Program Generation for Linear Algebra Using Multiple Layers of DSLs}

\author{
\IEEEauthorblockN{Daniele G. Spampinato\IEEEauthorrefmark{1},
Diego Fabregat-Traver\IEEEauthorrefmark{2}, Markus P\"uschel\IEEEauthorrefmark{1} and
Paolo Bientinesi\IEEEauthorrefmark{2}}\vspace{3mm}
\IEEEauthorblockA{\IEEEauthorrefmark{1}Department of Computer Science, ETH Zurich, Switzerland}
\IEEEauthorblockA{\IEEEauthorrefmark{2}Aachen Institute for Advanced Study in Computational Engineering Science, RWTH Aachen, Germany
\\[3mm]
Email: \IEEEauthorrefmark{1}\{danieles, pueschel\}@inf.ethz.ch, \IEEEauthorrefmark{2}\{fabregat, pauldj\}@aices.rwth-aachen.de
}}

\maketitle

\noindent
\mypar{Motivation}
%
Numerical software in computational science and engineering often relies on
highly-optimized building blocks from libraries such as BLAS~\cite{Dongarra:90}
and LAPACK~\cite{Lapack}.  Examples of such blocks include, but
are not limited to, matrix multiplications, matrix factorizations, and solvers
for Sylvester-like equations.
While the BLAS and LAPACK libraries have been very successful in providing
portable performance for a wide range of computing architectures, they
still present severe limitations in terms of flexibility.
%
%
First, these libraries are optimized for large matrices 
(of sizes at least in the hundreds). Second, the interface in terms of operations and matrix structures they provide 
specifically targets computational science.
These limitations can render those libraries suboptimal in performance or code size for applications in
communications, graphics, and control, which may require smaller scale computations
and a more flexible interface.

\mypar{Contributions}
To overcome these limitations, we advocate a domain-specific program
generator capable of producing library routines tailored to the specific needs of the
application in terms of sizes, interface, and target architecture.
In this work, we introduce such a generator that translates a desired linear algebra computation, annotated with matrix properties, into optimized C code, optionally vectorized with intrinsics. The generator unites prior work on two independent frameworks: 
The FLAME-based \click{}~\cite{Bientinesi:05,Fabregat:11a,
Fabregat:11b} and LGen~\cite{Spampinato:14,
Spampinato:16}, which was designed after Spiral~\cite{Pueschel:11}.

For a given linear algebra problem such as a matrix factorization, matrix inversion, or equation to be solved, 
\click{} synthesizes families of blocked algorithms that rely on basic computations provided by BLAS. These, in turn,
are compiled into efficient, vectorized C code by (an extension of) LGen.

Both tools use internally DSLs to perform the necessary reasonings and transformations at different levels of abstraction including
partitioning and traversal of operands and the preferred tiling shapes for locality and vectorization.
The integration of \click{} and LGen yields an end-to-end multi-DSL program generator that produces tailored library routines.

As case studies, we consider the Cholesky decomposition, and solvers for
the continuous-time Lyapunov and Sylvester equations. We compare the performance
of our generated code with the commercial Intel MKL showing competitive results.

\begin{figure}
\centering
\includegraphics[width=\columnwidth]{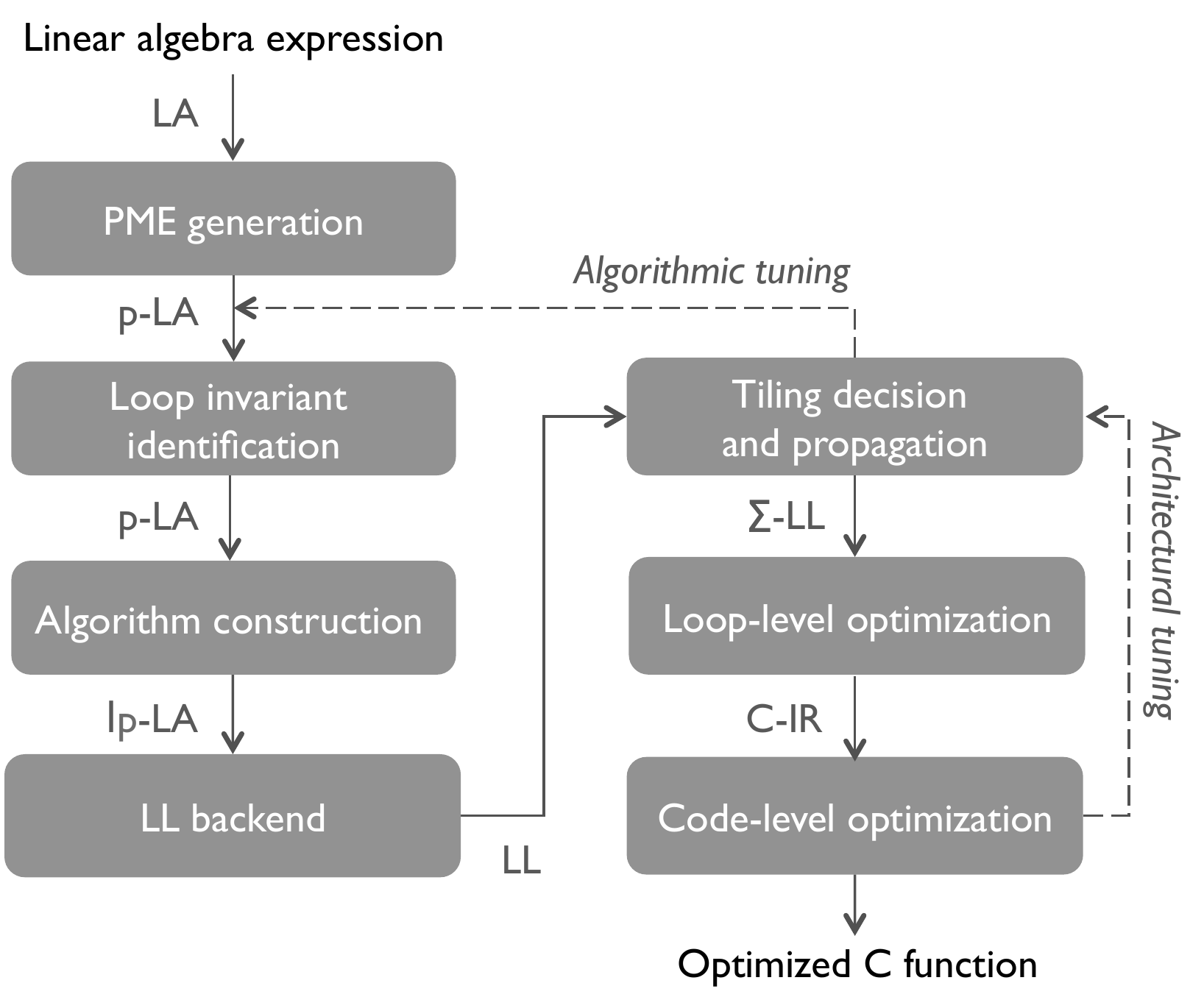}
\caption{Structure of our linear algebra generator. The left and right columns show
  processing stages at the algorithm and implementation level respectively.
  The arrow labels between boxes indicate DSLs at the input and output of each
  stage. All of DSLs are mathematical in nature, except for C-IR.
}
\label{fig:compiler}
\vspace{-4mm}
\end{figure}

\begin{figure*}[t]
\centering
\subfigure[]{\includegraphics[width=0.32\textwidth]{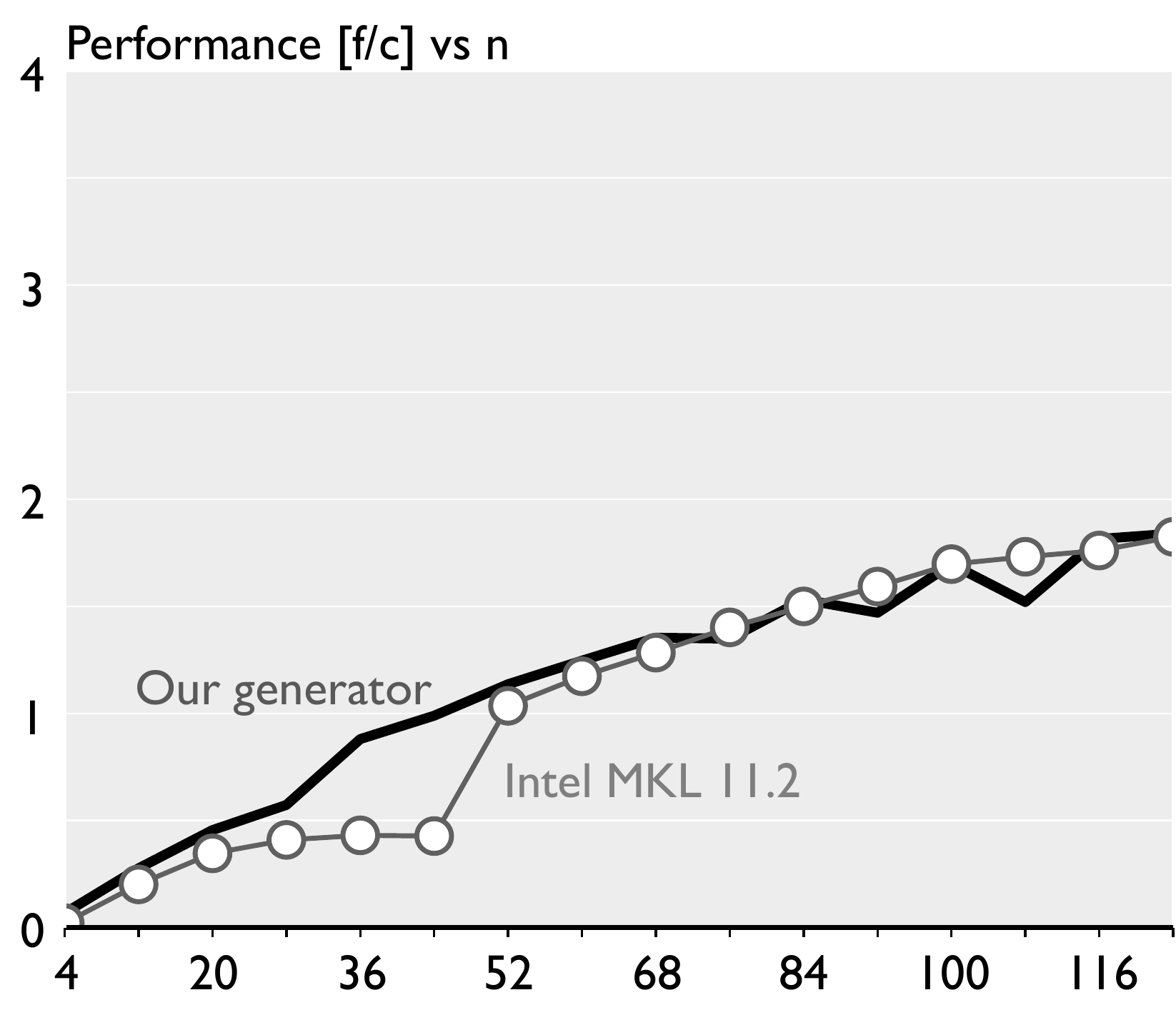}\label{fig:dupotrf}}
\subfigure[]{\includegraphics[width=0.32\textwidth]{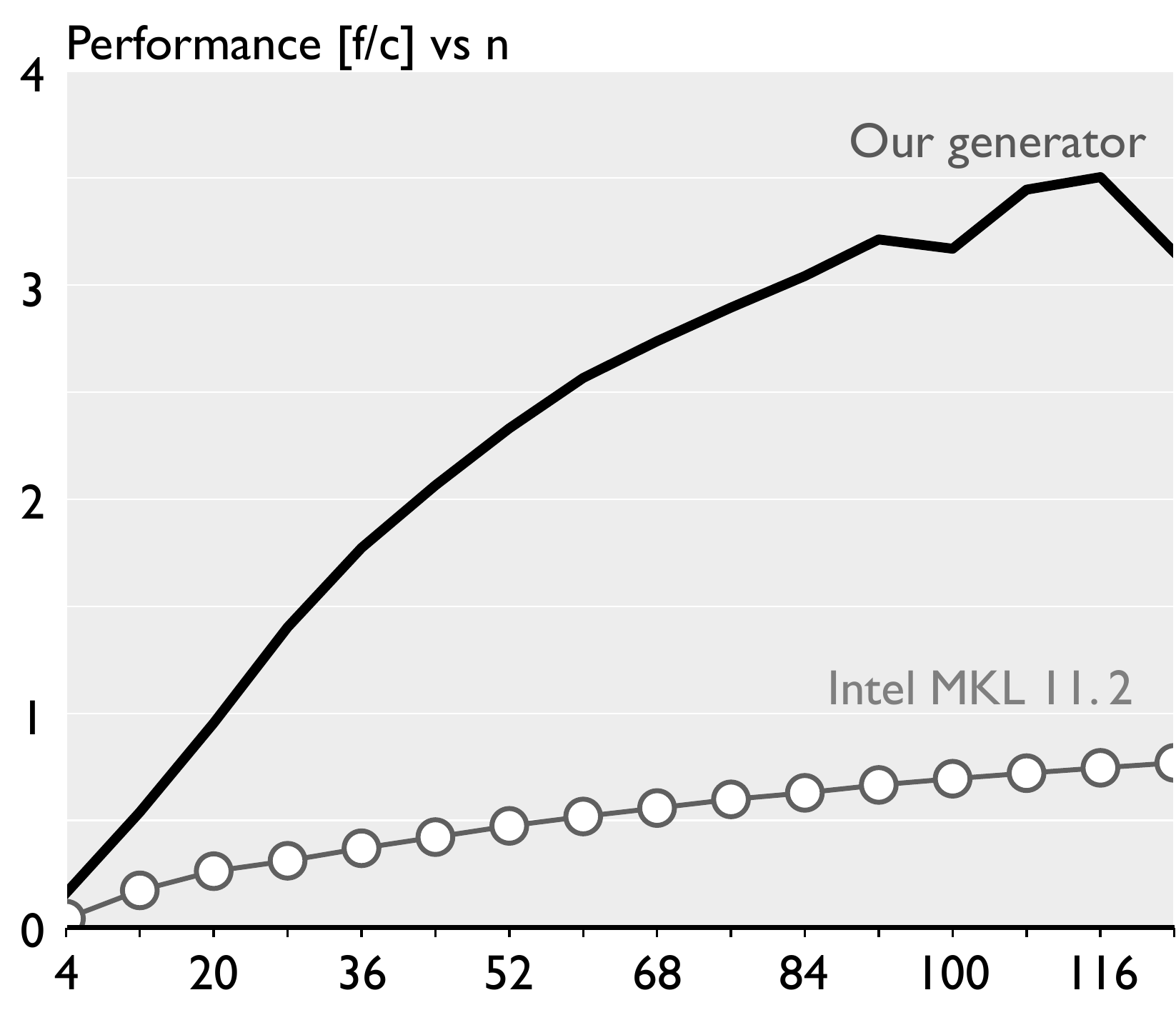}\label{fig:dllyap}}
\subfigure[]{\includegraphics[width=0.32\textwidth]{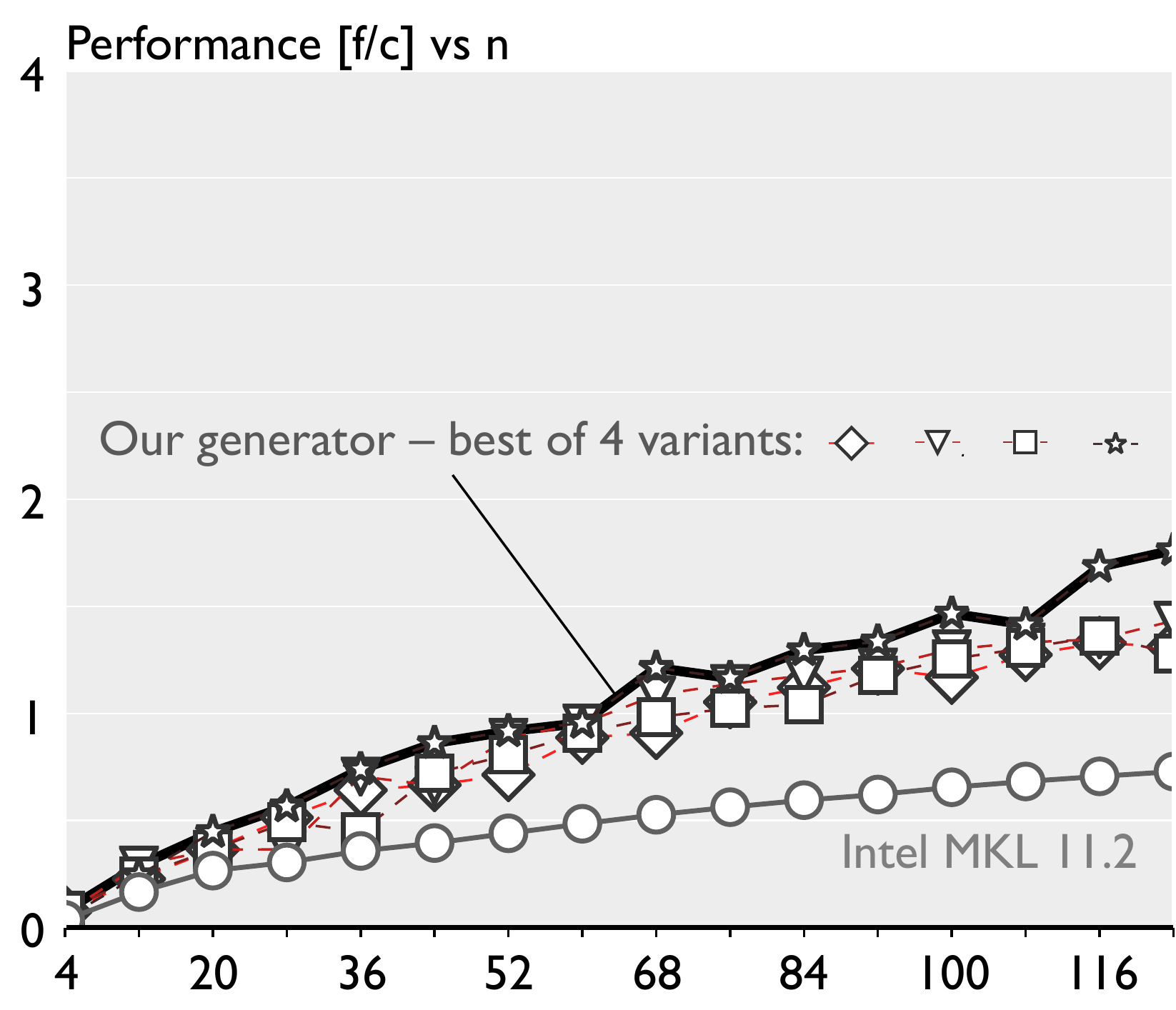}\label{fig:dlusylp}}
\caption{Performance results for: \subref{fig:dupotrf} $X_u^TX_u = A$,
    \subref{fig:dllyap} $L X_s + X_s L^T = S$, and \subref{fig:dlusylp} $LX +
    XU = C$. All matrices $\in\R{}^{n\times n}$; $A$, $L$, $S$, $U$, and $C$
    are inputs, $X_*$ are outputs; $L$ is lower triangular, $U$, $X_u$ are
    upper triangular, $S$, $X_s$ are symmetric, and $A$ is symmetric positive
    definite.  In~\subref{fig:dupotrf} $f \approx \frac{n^3}{3}$ flops while in
    \subref{fig:dllyap}--\subref{fig:dlusylp} $f \approx 2n^3$ flops.
    Tests compiled with icc v.16 
    and run on an Intel Sandy Bridge (AVX, 32 kB L1-D cache, 256 kB
    L2 cache) under Ubuntu 14.04 (Linux v.3.13). Performance in
    flops per cycle (f/c) vs. $n$ (order of the matrices). Theoretical peak performance of the processor
    is 8 f/c in all cases (assuming balanced additions and multiplications with perfect vectorization efficiency)
    but we scale to 4 f/c for better readability.}
\label{fig:plots}
\vspace{-4mm}
\end{figure*}

\mypar{Generator structure}
Fig.~\ref{fig:compiler} illustrates the structure of our linear algebra generator.
The input is an annotated linear algebra expression, e.g., $X^TX=A$ for a Cholesky decomposition, where $A$ is the input matrix, symmetric, positive definite, and $X$ is the output matrix and upper triangular. The output of the generator is an optimized C function, optionally vectorized with intrinsics.

The generator integrates the processing stages from \click{} (left column in Fig.~\ref{fig:compiler}) and
LGen (right column) into a single framework. In particular, the stages on the left take the input expression and
synthesize for it several algorithms. To do so,
first the input equation is transformed into one or more Partitioned
Matrix Expressions (PMEs), \ie{}, a recursive definition of the equation;
then, the PMEs are decomposed to identify loop invariants;
finally, a family of loop-based algorithms are built around these loop
invariants using formal methods techniques.  
The synthesized algorithms are mathematically equivalent, but may
yield different performance.

Given a particular algorithm, the stages on the right in Fig.~\ref{fig:compiler} proceed with its implementation.
In particular, matrices are tiled for locality and summation-based definitions of each
statement in the algorithm is determined taking matrix structures into account. 
Decisions at this stage are either search-based (\eg{}, tile sizes) or model-based (\eg{}, matrices traversal). 
Vectorization requires an additional level of tiling to expose operations whose sizes match the vector length of the CPU (see $\nu$-BLACs in~\cite{Spampinato:14}). These are later mapped to a small set of pre-implemented 
basic blocks efficiently vectorized with intrinsics.

Finally, autotuning (dashed arrows in Fig.~\ref{fig:compiler}) is used to select
the fastest code searching among alternative algorithms and implementation
parameters.

\mypar{A multi-DSL approach}
All processing stages in Fig.~\ref{fig:compiler} but the last one (code-level
optimization) operate on mathematical DSLs at different levels of abstraction.
More specifically, the LA language is a subset of standard linear algebra notation.
The p-LA (partitioned LA) language is a refinement of LA which represents
equations in terms of partitioned operands (submatrices and subvectors)
together with knowledge inferred during the partitioning of the operands.
The lp-LA (loop-based p-LA) notation introduces programming constructs such as
explicit loops, and encodes algorithms at a high-level of abstraction,
resembling the FLAME notation~\cite{Bientinesi:05}.  

The LL (Linear algebra Language) and \sll{} DSLs represent basic linear algebra operations but at a lower level of abstraction.
They include explicit gather and scatter operators on matrices and capture matrix structures
mathematically using techniques from polyhedral compilation~\cite{Bastoul:04}.
The LL backend stage (bottom-left in Fig.~\ref{fig:compiler}) is used to bridge
lp-LA and LL notations. 

\mypar{Preliminary performance results}
In Fig.~\ref{fig:plots} we show performance results for three problems: \subref{fig:dupotrf} the Cholesky decomposition, 
and the solution of triangular, continuous-time \subref{fig:dllyap} Lyapunov and \subref{fig:dlusylp} Sylvester equations. 
Computations are performed in double precision, and all matrices
are square with dimensions divisible by four (the vector length of AVX). We compare the AVX-vectorized routines generated by our
system specific for each input size with the corresponding routines from the LAPACK implementation in Intel MKL (which is generic in size).

For each case, our generator explores the implementation of different algorithms. 
For example, in Fig.~\ref{fig:dlusylp} the final code is chosen among four different variants. 
The specialized code produced by our generator can be as fast as MKL (Fig.~\ref{fig:dupotrf}) or outperforms it (between $4\times$ and 
$4.8\times$ in Fig.~\ref{fig:dllyap}, and between $2\times$ and $2.5\times$ in Fig.~\ref{fig:dlusylp}).


\mypar{Limitations and future work}
We presented ongoing work and are currently addressing a few limitations of our generator. In particular: (1) the number
of temporary submatrices in the algorithm construction phase can be reduced; (2) more complete and efficient vectorization including for sizes not divisible by the vector length.

Finally, we are considering the replacement of the current autotuning loops with a model-based approach.

\bibliographystyle{abbrvnat}

\begin{thebibliography}{26}
\footnotesize
\bibitem[Dongarra et~al.(1990)Dongarra, Du~Croz, Hammarling, and
  Duff]{Dongarra:90}
J.~J. Dongarra {\it et al}.
\newblock A set of level 3 basic linear algebra subprograms.
\newblock \emph{ACM Trans. on Mathematical Software (TOMS)}, 16\penalty0
  (1):\penalty0 1--17, 1990.

\bibitem[Anderson et~al.(1999)Anderson, Bai, Bischof, Blackford, Demmel,
  Dongarra, Du~Croz, Greenbaum, Hammarling, McKenney, and Sorensen]{Lapack}
E.~Anderson {\it et al}.
\newblock \emph{{LAPACK} Users' Guide}.
\newblock Society for Industrial and Applied Mathematics, third edition, 1999.

\bibitem[P.~Bientinesi et~al.(2005)]{Bientinesi:05}
P. Bientinesi, J. A. Gunnels, M. E. Myers, E. S. Quintana-Ort\'{i}, and R. A. van~de~Geijn.
\newblock The science of deriving dense linear algebra algorithms.
\newblock \emph{ACM Trans. on Mathematical Software (TOMS)}, 31\penalty0
  (1):\penalty0 1--26, 2005.


\bibitem[Fabregat and Bientinesi(2011a)]{Fabregat:11a}
D.~Fabregat-Traver and P.~Bientinesi.
\newblock Automatic Generation of Loop-Invariants for Matrix Operations.
\newblock In \emph{Computational Science and Its Applications
  (ICCSA)}, pp. 82--92, 2011.

\bibitem[Fabregat and Bientinesi(2011b)]{Fabregat:11b}
D.~Fabregat-Traver and P.~Bientinesi.
\newblock Knowledge-Based Automatic Generation of Partitioned Matrix Expressions.
\newblock In \emph{Computer Algebra in Scientific Computing (CASC)}, vol. 6885 of \emph{Lecture
 Notes in Computer Science (LNCS)}, pp. 144--157. Springer, 2011.

\bibitem[P{\"u}schel et~al.(2011)P{\"u}schel, Franchetti, and
  Voronenko]{Pueschel:11}
M.~P{\"u}schel, F.~Franchetti, and Y.~Voronenko.
\newblock \emph{Encyclopedia of Parallel Computing}, chap. Spiral.
\newblock Springer, 2011.

\bibitem[Spampinato and P{\"u}schel(2014)]{Spampinato:14}
D.~G. Spampinato and M.~P{\"u}schel.
\newblock A basic linear algebra compiler.
\newblock In \emph{Code Generation and Optimization
  (CGO)}, pp. 23--32, 2014.

\bibitem[Spampinato and P{\"u}schel(2016)]{Spampinato:16}
D.~G. Spampinato and M.~P{\"u}schel.
\newblock A basic linear algebra compiler for structured matrices.
\newblock In \emph{Code Generation and Optimization
  (CGO)}, pp. 117--127, 2016.



\bibitem[Bastoul(2004)]{Bastoul:04}
C.~Bastoul.
\newblock Code generation in the polyhedral model is easier than you think.
\newblock In \emph{Parallel Architectures and Compilation Techniques (PACT)},
  pp. 7--16, 2004.


\end{thebibliography}

\end{document}